\documentclass[pre,twocolumn]{revtex4}
\usepackage{epsfig}
\usepackage{amsmath}
\usepackage{hyperref}
\usepackage{breakurl}

\begin{document}

\title{Passive scalar tracers in chaotic and turbulent atmosphere}

\author{A. Bershadskii}

\affiliation{
ICAR, P.O. Box 31155, Jerusalem 91000, Israel}

\begin{abstract}
  The notion of helical distributed chaos has been used for the description of the spatio-temporal dynamics of the passive scalar tracers from the atmospheric surface layer up to the mesosphere (at different stability conditions and different latitudes). Point, linear and distributed sources and sinks with different types of artificial and natural trace-gases (ozone, carbon dioxide, etc.), and aerosols have been studied. Related properties of the solar radiation transmittance in the cloudy atmosphere have been also briefly discussed as well as analogous properties of the tracers in Jupiter's atmosphere.

\end{abstract}

\maketitle

\section{Deterministic chaos}

\begin{figure} \vspace{-0.5cm}\centering
\epsfig{width=.45\textwidth,file=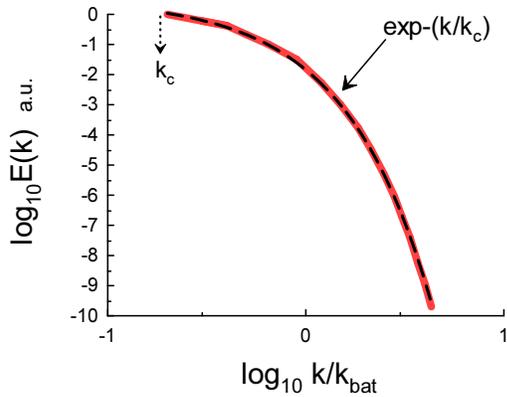} \vspace{-4cm}
\caption{Power spectrum of passive scalar fluctuations for isotropic homogeneous (steady) spatio-temporal fluid chaos ($Re_{\lambda} = 8$ and $Sc = 1$).}
\end{figure}

  It is known that the chaotic smooth dynamical systems with compact strange attractors usually have exponential {\it frequency} spectra \cite{oh}-\cite{mm}.  For the systems described by the equations with partial derivatives, one can expect that the spatio-temporal smoothness will also result in the wavenumber exponential spectra.\\
 
 The dynamics of a passive scalar tracer $\chi$ in a velocity field ${\bf u}$ can be described by equation
$$
\partial_t \chi+ \bf{u}\cdot\nabla\chi= \kappa\nabla^2\chi+f_{\chi}  \eqno{(1)}
$$

   Results of a direct numerical simulation of the passive scalar tracers mixing in the statistically stationary isotropic homogeneous fluid motion 
described  by the incompressible Navier-Stokes equations
$$  
  \partial_t {\bf u} + {\bf u}\cdot \nabla {\bf u} = -\nabla p +\nu \nabla^2 {\bf v}+\mathbf{f}, \eqno{(2)}
$$
$$  
  \nabla \cdot \bf{u}=0   \eqno{(3)}
$$  
for small Reynolds numbers were reported in the paper Ref. \cite{dsy}. \\

  Figure 1 shows the wavenumber power spectra for such passive scalar tracer mixing for the Taylor-Reynolds number $Re_{\lambda}=8$ (the Schmidt number $Sc=1$ and the periodic boundary conditions were used). For such a small Reynolds number the passive scalar concentration field $\chi$ can be represented by deterministic chaos and has an exponential wavenumber power spectrum
  
 $$
 E(k) \propto \exp-(k/k_c) \eqno{(4)}
$$ 

The position of the characteristic wavenumber $k_c$ is indicated in the Fig. 1 by the dotted arrow, $k_{bat}$ is the Batchelor wavenumber and the dashed curve is drawn to indicate the exponential type of the spectrum. The spectral data for the Fig. 1 were taken from Fig. 1a of the Ref. \cite{dsy} (see the Ref. \cite{dsy} for a more detail description of the numerical simulation).

\section{Buoyancy-driven thermal convection}

  The buoyancy-driven thermal convection in the Boussinesq approximation can be described by equations \cite{kcv}
$$
\frac{\partial {\bf u}}{\partial t} + ({\bf u} \cdot \nabla) {\bf u}  =  -\frac{\nabla p}{\rho_0} + \sigma g \theta {\bf e}_z + \nu \nabla^2 {\bf u}   \eqno{(5)}
$$
$$
\frac{\partial \theta}{\partial t} + ({\bf u} \cdot \nabla) \theta  =  S  \frac{\Delta}{H}e_z u_z + \kappa \nabla^2 \theta, \eqno{(6)}
$$
$$
\nabla \cdot \bf u =  0 \eqno{(7)}
$$
where $\theta$ is the temperature fluctuations over an imposed temperature profile $T_0 (z)$, ${\bf e}_z$ is a unit vector along with the gravity force, $g$ is the gravity acceleration, $\Delta$ is the temperature difference between the different  layers, $H$ is the distance between the layers, $\rho_0$ is the mean density, $\sigma$ is the thermal expansion coefficient, $\nu$ and $\kappa$ are the viscosity  and thermal diffusivity. For the unstable stratification the parameter $S=+1$ and for the stable stratification it is $S=-1$.\\

 For the inviscid case (when $Ra \rightarrow \infty$) the equation for the mean helicity is
$$
\frac{d\langle h \rangle}{dt}  = 2\langle {\boldsymbol \omega}\cdot {\bf F}  \rangle \eqno{(8)} 
$$ 
with
$$
{\bf F} = (\sigma g \theta) {\bf e}_z,  \eqno{(9)}
$$
here ${\boldsymbol \omega} = \nabla \times {\bf u}$ and $h={\bf u}\cdot {\boldsymbol \omega}$ are the vorticity and helicity fields respectively,  and $\langle...\rangle$ is an average over the liquid volume. It is clear from the Eq. (8) that the mean helicity is not an inviscid invariant for the  buoyancy-driven thermal convection.  We will consider, therefore, the case when the large-scale motions only will contribute the main part to the correlation $\langle {\boldsymbol \omega}\cdot {\bf F}  \rangle$. The correlation $\langle {\boldsymbol \omega}\cdot {\bf F}  \rangle$, however, is rapidly decreasing with spatial scales (such situation is typical for the highly chaotic flows). Therefore, in spite of the mean helicity is not an  inviscid invariant here the higher moments of the helicity distribution $h={\bf u}\cdot {\boldsymbol \omega}$ can be still (approximately) considered as the inviscid invariants \cite{lt},\cite{mt}.\\

\begin{figure} \vspace{-1.6cm}\centering
\epsfig{width=.47\textwidth,file=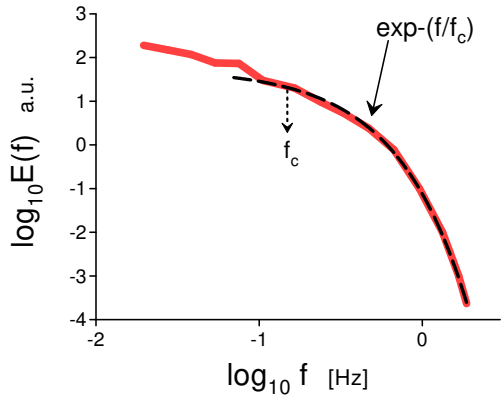} \vspace{-3.6cm}
\caption{Power spectrum of the concentration fluctuations for the passive scalar tracers ejected from a line source into the atmospheric surface layer at stable conditions.}
\end{figure}
\begin{figure} \vspace{-1.5cm}\centering
\epsfig{width=.47\textwidth,file=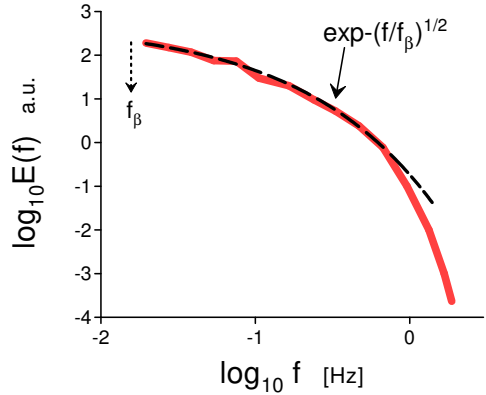} \vspace{-3.7cm}
\caption{As in the Fig. 2 but the dashed curve indicates the stretched exponential spectral law Eq. (15) for the small $f$.}
\end{figure}
\begin{figure} \vspace{-0.5cm}\centering
\epsfig{width=.47\textwidth,file=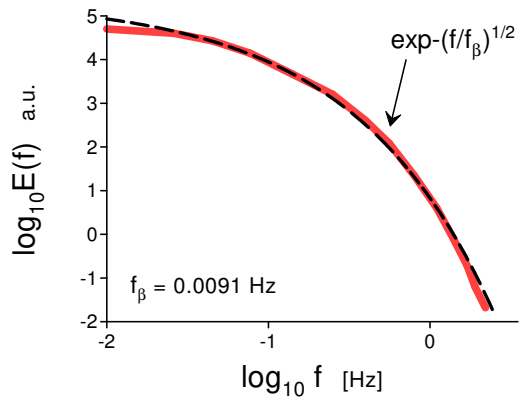} \vspace{-3.8cm}
\caption{Power spectrum of the concentration fluctuations for the passive scalar tracers ejected from a linear source in the atmospheric surface layer over a homogeneous canopy at mildly unstable to very unstable conditions.}
\end{figure}

   To show this, we will divide the liquid volume into a net of the cells which are moving with the liquid (the Lagrangian description) - $V_i$ \cite{lt}\cite{mt}. The boundary conditions on their surfaces are taken as ${\boldsymbol \omega} \cdot {\bf n}=0$. Moments of order $n$ can be defined as 
 $$
I_n = \lim_{V \rightarrow  \infty} \frac{1}{V} \sum_j H_{j}^n  \eqno{(10)}
$$
where the total helicity $H_j$ in the subvolume $V_j$ is
$$
H_j = \int_{V_j} h({\bf r},t) ~ d{\bf r}.  \eqno{(11)}
$$
   Due to the fast reduction of the correlation $\langle {\boldsymbol \omega}\cdot {\bf F} \rangle$ with the scales the helicities $H_j$ can be approximately considered as inviscid invariants for the cells with the small enough characteristic spatial scales. Just these cells provide the main contribution to the moments  $I_n$ with $n \gg 1 $ for the chaotic flows (cf. \cite{bt}).  Therefore, the  moments $I_n$ for the sufficiently large $n$ can be considered as inviscid quasi-invariants while the total helicity $I_1$ cannot. For the strongly chaotic flows the values $n=2$ and $n=3$ can be still considered as sufficiently large  (where the moment $I_2$ is the so-called Levich-Tsinober invariant of the Euler equation \cite{lt}). For the viscous cases such moments can be considered as {\it adiabatic} invariants in the inertial range of scales. \\ 
   
\begin{figure} \vspace{-1.5cm}\centering
\epsfig{width=.47\textwidth,file=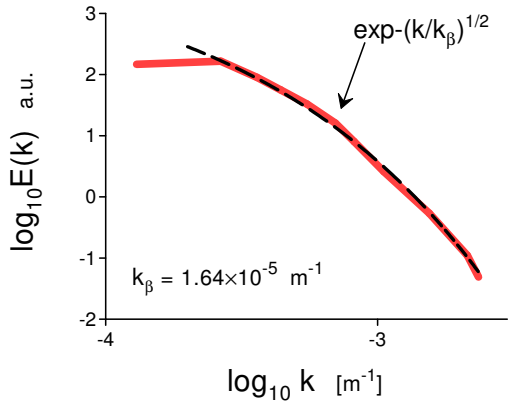} \vspace{-3.9cm}
\caption{Power spectrum of the ozone mixing ratio fluctuations in the lower stratosphere.}
\end{figure}
 
\begin{figure} \vspace{-0.3cm}\centering
\epsfig{width=.47\textwidth,file=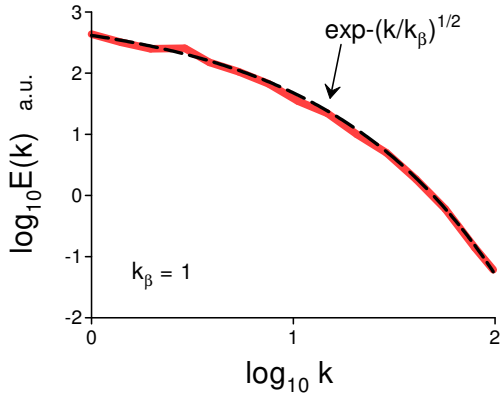} \vspace{-4cm}
\caption{Power spectrum of the ozone mixing ratio fluctuations in the global upper troposphere and lower stratosphere.}
\end{figure}
\begin{figure} \vspace{-0.5cm}\centering
\epsfig{width=.47\textwidth,file=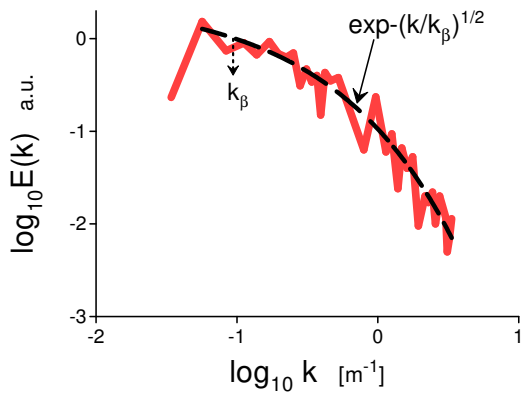} \vspace{-4cm}
\caption{ Power spectrum of the aerosols concentration fluctuations measured in situ in the upper mesosphere.}
\end{figure}
   The chaotic attractors (in the phase space) correspond to the adiabatic invariants $I_n$. Their basins of attraction, however, can be significantly different: the attractors corresponding to the larger $n$ have a thinner basin of attraction (the intermittency). Therefore, the flow dynamics is dominated by the invariant $I_n$ with the smallest (available) order $n$. \\
   
   We will begin consideration from  $I_3$, for simplicity. The dimensional considerations can be used to estimate characteristic velocity $u_c$ for the fluctuating $k_c$ 
 $$
 u_c \propto |I_3|^{1/6} k_c^{1/2}    \eqno{(12)}
 $$
in this case

\section{Helical distributed chaos}

     Figure 2 shows power spectrum of the concentration fluctuations for the passive scalar tracers (sulfur hexafluoride - $SF_6$) ejected from a line source into the atmospheric surface layer at stable conditions. The spectral data were taken from Fig. 1 of the Ref. \cite{finn}. This is a frequency ($E(f)$) spectrum obtained from the time series of the passive scalar concentration measured by a probe. The frequency spectrum represents actually the spatial structures passing through the probe with the mean velocity of the wind $U_0$ and the frequency spectrum should be converted into a wavenumber spectrum using the Taylor's `frozen' turbulence hypothesis \cite{my} with $f = U_0~k/2\pi$, where $k$ is the wavenumber. Therefore, the dashed curve is drawn in the Fig. 2 to indicate the exponential spectral law Eq. (4). This spectral law gives a good approximation for the large frequencies, i.e. for the large wavenumbers where the diffusivity (viscosity) at the moderately stable conditions provides existence of a deterministic chaos. For the small wavenumbers, however, the influence of the diffusive processes is not strong enough and the parameter $k_c$ in the Eq. (4) becomes fluctuating. These fluctuations can be taken into account with an ensemble average 
$$
E(k) \propto \int_0^{\infty} P(k_c) \exp -(k/k_c)~ dk_c  \eqno{(13)}
$$
 here $P(k_c)$ is the probability distribution of the characteristic scale $k_c$. \\
 
    Assuming a Gaussian (normal) distribution of the characteristic velocity $u_c$  \cite{my}, one can obtain the distribution $P(k_c)$ for the helically dominated distributed chaos from the Eq. (12), for instance,
$$
P(k_c) \propto k_c^{-1/2} \exp-(k_c/4k_{\beta})  \eqno{(14)}
$$
where the new parameter $k_{\beta}$ is a constant (see below).

    Substituting the Eq. (14) into the Eq. (13) one obtains
$$
E(k) \propto \exp-(k/k_{\beta})^{1/2}  \eqno{(15)}
$$     
 
 for the helically dominated distributed chaos.\\
 
   Figure 3 shows power spectrum of the concentration fluctuations for the passive scalar tracers as it is shown in the Fig. 2 but now the dashed curve indicates the stretched exponential spectral law Eq. (15) for the small values of the wavenumber $k$.\\
   
   Figure 4 shows power spectrum of the concentration fluctuations for the passive scalar tracers over a homogeneous canopy (a peach orchard)
of intermediate roughness ejected from a line source in the atmospheric surface layer at mildly unstable to very unstable conditions. The spectral data were taken from Fig. 2b of the Ref. \cite{lmf} (the spectrum was computed using Yule-Walker AR method). The dashed curve indicates the stretched exponential spectral law Eq. (15) \\

\begin{figure} \vspace{-1.5cm}\centering
\epsfig{width=.45\textwidth,file=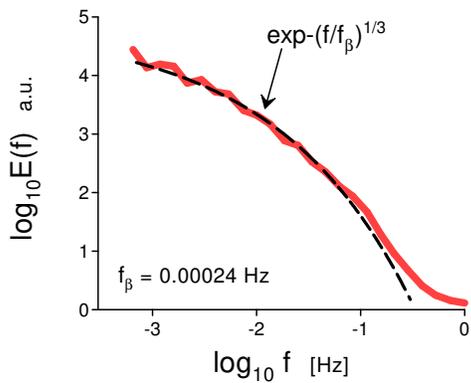} \vspace{-3.9cm}
\caption{ Power spectrum of the of the naturally emitted nitric oxide (NO) within an Amazonian rain forest at the height 1m above the forest floor.}
\end{figure}
\begin{figure} \vspace{-0.5cm}\centering
\epsfig{width=.45\textwidth,file=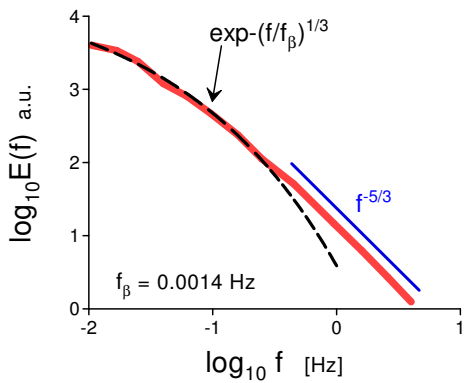} \vspace{-3.9cm}
\caption{ Power spectrum of the CO$_2$ concentration fluctuations measured in the atmospheric surface layer over an abandoned peat meadow with comparatively high CO$_2$ uptake. }
\end{figure}

    The results of the 4 year measurements made with an ozonesonde in the lower stratosphere over a latitudinal region 60$^o$S-43$^o$N were reported in the paper Ref. \cite{ogi}. A research oceanic vessel was used as a main base. The measurements were made in the altitude range 19-27 km. \\
    
     Figure 5 shows power spectrum of the ozone mixing ratio fluctuations (averaged over all measurements). The spectral data were taken from the Fig. 5 of the Ref. \cite{ogi}. The dashed curve is drawn to indicate the stretched exponential spectrum Eq. (15).\\
     
     Figure 6 shows power spectrum of the ozone distribution obtained at 55$^o$N in February 2005 (the spectral data were taken from Fig. 4b of the paper Ref. \cite{warg}). The assimilation in the upper troposphere and lower stratosphere was studied with the NASA’s EOS‐Aura ozone data using meteorological fields from the Goddard Earth Observing System. The dashed curve is drawn to indicate the stretched exponential spectrum Eq. (15). The wavenumber $k_{\beta} =1$ corresponds to the longest planetary wave and, as it follows from the Fig. 6, the entire helical distributed chaos is tuned to the longest planetary wave.\\
   
     The ice particles in the upper mesosphere are surrounded by ionospheric plasma. Therefore, they gain a charge because of ion and electron capture processes. Figure 7 shows the power spectrum of the charged particles (aerosols) concentration measured in situ in the upper mesosphere at the height between 85.3 - 85.5 km during the ECOMA/MASS rocket campaign at 69$^o$N in August 2007. The spectral data were taken from the Fig. 6 of the Ref. \cite{strel}. The charged particles were highly structured at these heights.The estimated Schmidt number $Sc = 541$. The dashed curve is drawn to indicate the stretched exponential spectrum Eq. (15). \\

 \section{Strong distributed chaos}
 
  It is not surprising that for the {\it smooth} dynamical systems the spectrum has the stretched exponential form. The particular stretched exponential Eq. (15) can be generalized
$$
E(k) \propto \exp-(k/k_{\beta})^{\beta},   \eqno{(16)}
$$ 
where the $k_{\beta}$ is a constant. 
  
   In order to find value of the $\beta$ we can use asymptotic (at $k_c \rightarrow \infty$) properties of the distribution $P(k_c)$.  It follows from the Eqs. (13) and (16) at this asymptotic \cite{jon}
$$
P(k_c) \propto k_c^{-1 + \beta/[2(1-\beta)]}~\exp(-bk_c^{\beta/(1-\beta)}), \eqno{(17)}
$$
with $b$ as a constant. On the other hand, the asymptotic distribution $P(k_c)$ can be found from the dimensional considerations. Indeed, the particular estimate Eq. (12) can be replaced by a general estimate
$$
 u_c \propto |I_n|^{1/2n}~ k_c^{\alpha_n}    \eqno{(18)}
 $$    
where 
$$
\alpha_n = 1-\frac{3}{2n}  \eqno{(19)}
$$  

 If again $u_c$ has Gaussian (normal) distribution a relationship between the exponents $\beta_n$ and $\alpha_n$ can be readily obtained from the Eqs. (17) and (18)
$$
\beta_n = \frac{2\alpha_n}{1+2\alpha_n}  \eqno{(20)}
$$
 
  Substituting $\alpha_n $  from the Eq. (19) into the Eq. (20) we obtain
 $$
 \beta_n = \frac{2n-3}{3n-3}   \eqno{(21)}  
 $$
 
  For strongly chaotic helicity field even value $n=2$ can be considered as sufficiently large to consider the second moment $I_2$ (the Levich-Tsinober invariant of the Euler equation \cite{lt},\cite{mt}) as an adiabatic invariant (see Section III). In this case we obtain from the Eq. (21) $\beta =1/3$, i.e.
$$
E(k) \propto \exp-(k/k_{\beta})^{1/3}  \eqno{(22)}
$$ 
\begin{figure} \vspace{-1.5cm}\centering
\epsfig{width=.45\textwidth,file=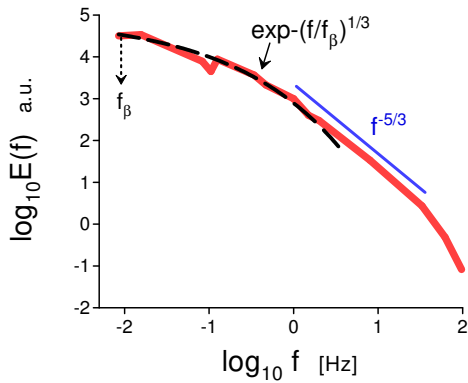} \vspace{-3.9cm}
\caption{ Power spectrum of a passive scalar tracer ($C_3H_6$) in a dispersing plume artificially ejected from an elevated point source in the atmospheric surface layer over a flat and smooth area at near-neutral stability conditions. }
\end{figure}
\begin{figure} \vspace{-1.6cm}\centering
\epsfig{width=.45\textwidth,file=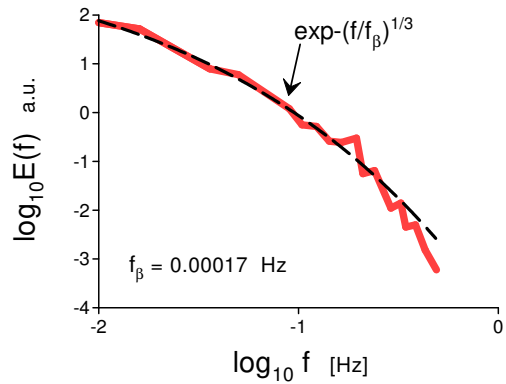} \vspace{-3.8cm}
\caption{ Power spectrum of the ozone concentration fluctuations in the anvil of a large cumulonimbus cloud (of a severe thunderstorm). }
\end{figure}
\begin{figure} \vspace{-0.5cm}\centering
\epsfig{width=.45\textwidth,file=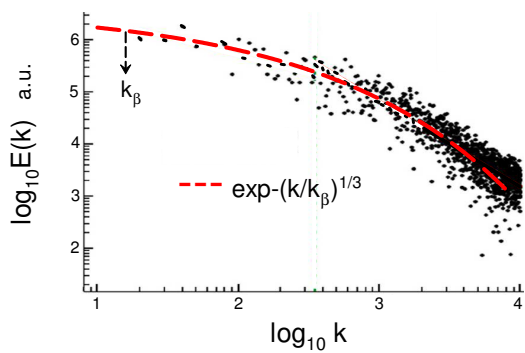} \vspace{-4.1cm}
\caption{Spectrum of the intensity fluctuations of the Jupiter's upper cloud layer reflectivity. } 
\end{figure}

  Figure 8 shows power spectrum of the naturally emitted nitric oxide (NO) within an Amazonian rain forest at the height 1m above the forest floor. The spectral data were taken from the Fig. 3 of the Ref. \cite{rumm}. The dashed curve indicates the stretched exponential spectral law Eq. (22).\\
  
  The wet peat lands are considered as a sink of CO$_2$. Figure 9 shows power spectrum of the CO$_2$ concentration fluctuations measured over an abandoned peat meadow in the Netherlands with comparatively high CO$_2$ uptake (the spectral data were taken from the Fig. 2 of the Ref.\cite{hend}). The measurement were performed in the atmospheric surface layer. The dashed curve indicates the stretched exponential spectral law Eq. (22).\\
  
  Figure 10 shows power spectrum of a passive scalar tracer ($C_3H_6$) in a dispersing plume artificially ejected from an elevated point source in the atmospheric surface layer (at a release height of 2.5 m). The experimental site was flat and smooth. The passive scalar tracer was released isotropically
(without forming a jet). The measurements were performed near the mean-plume centerline under near-neutral stability conditions (the spectral data were taken from the Fig. 19 of the Ref. \cite{yee}). The dashed curve indicates the stretched exponential spectral law Eq. (22).\\

\begin{figure} \vspace{-1.2cm}\centering
\epsfig{width=.49\textwidth,file=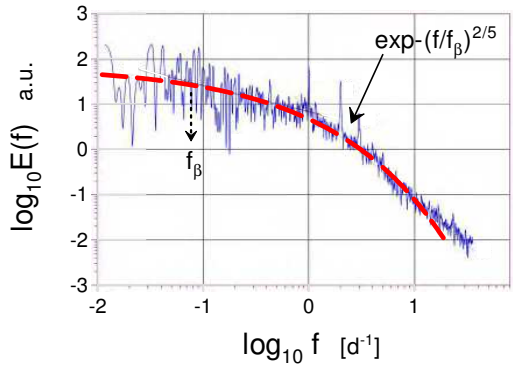} \vspace{-5.2cm}
\caption{Power spectrum of the  the atmospheric $NO_2$ temporal fluctuations in the center of the Moscow megacity.} 
\end{figure}
\begin{figure} \vspace{-0.5cm}\centering
\epsfig{width=.45\textwidth,file=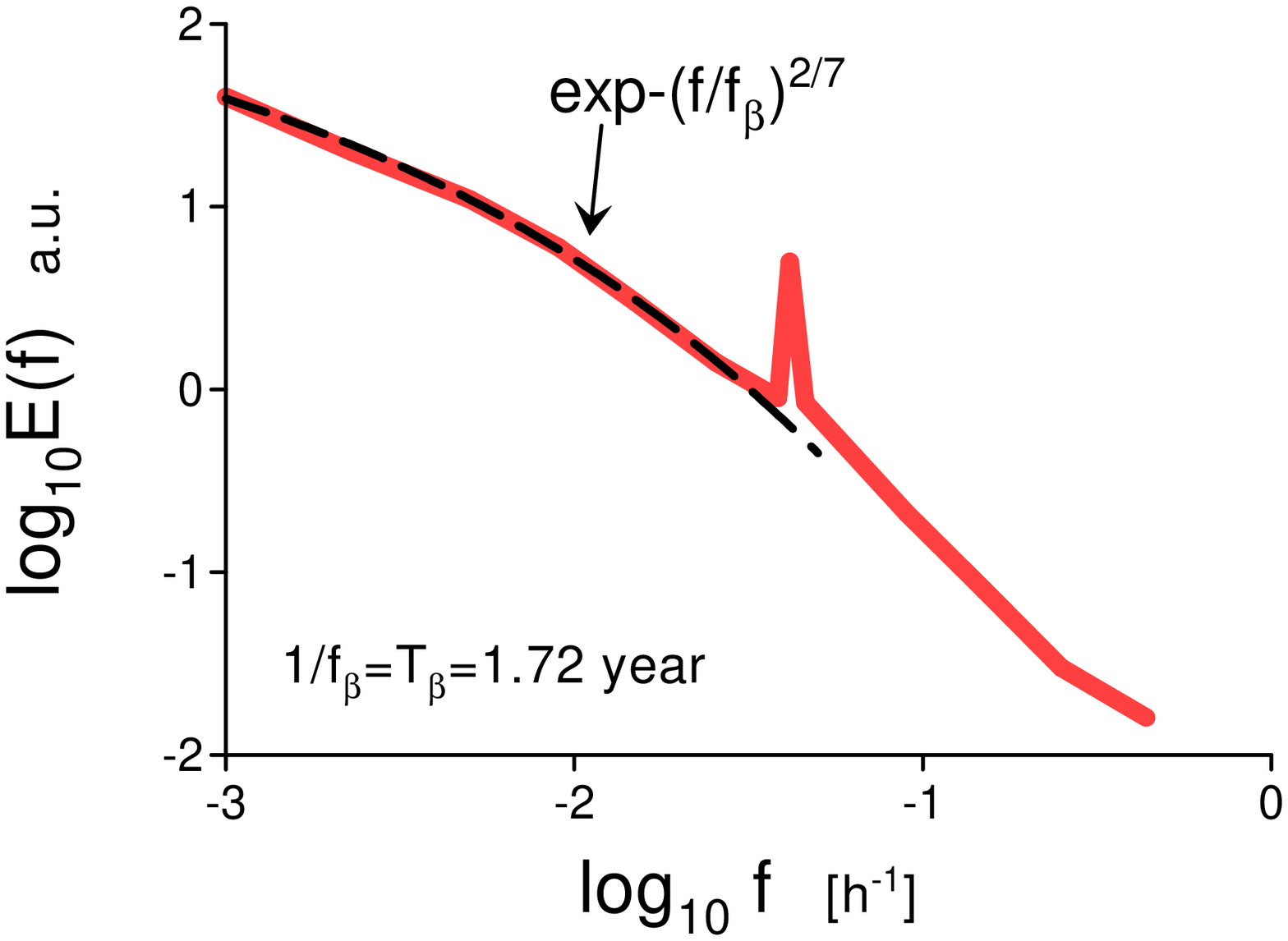} \vspace{-3.9cm}
\caption{Power spectrum of the atmospheric $CO_2$ temporal fluctuations - high latitude (northern hemisphere). } 
\end{figure}
\begin{figure} \vspace{-1cm}\centering
\epsfig{width=.45\textwidth,file=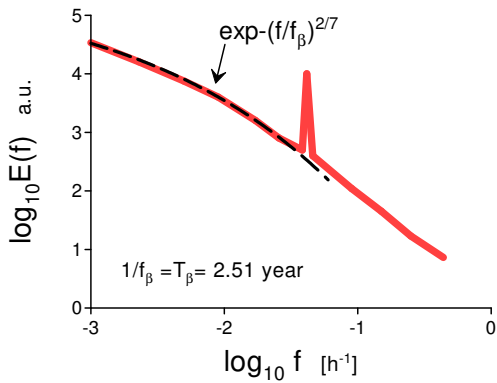} \vspace{-3.9cm}
\caption{Power spectrum of the atmospheric $CO_2$ temporal fluctuations - middle latitude (northern hemisphere). } 
\end{figure}

  Figure 11 shows power spectrum of the ozone concentration fluctuations in the anvil of a large cumulonimbus cloud (of a severe thunderstorm). The spectral (aircraft-measured) data were taken from the Fig. 7 of the Ref. \cite{dh}. The measurements were performed in the convectively neutral region of the anvil, corresponding approximately to the altitude of the midsection of the thunderstorm core. The dashed curve indicates the stretched exponential spectral law Eq. (22).\\
  
\section{Jupiter's upper cloud layer}
  
   The Jupiter's upper atmosphere is characterized by banded continual appearance of the dark and light clouds (see, for instance, Ref. \cite{for} and references therein). Complete maps of Jupiter's upper cloud layer obtained using the Hubble Space Telescope (at optical wavelengths) were reported in recent Ref. \cite{csm}.  The observed global Jupiter's maps were then used in the Ref. \cite{csm} in order to compute corresponding passive scalar trasers' power spectra for single rotation observations. An example of the spectrum of the intensity fluctuations of the cloud's reflectivity is plotted in the Fig. 12 against the longitudinal wave number $k$ (which represents the number of complete sinusoids fitting in the circumference of the planet at a chosen latitude). The spectral data were taken from Fig. 3 of the Ref. \cite{csm}. The dashed curve in the Fig. 12 indicates correspondence to the spectrum Eq. (22).

\section{Temporal distributed chaos}

   In the previous sections the spatial distributed chaos has been studied and the observed frequency spectra were a direct reflection of the spatial chaotic processes due to the Taylor's `'frozen'' fluctuations mechanism. In order to observe real temporal fluctuations one needs in measurements performed for larger temporal scales (see below).  \\
   
    Let us consider the temporal distributed chaos.  In this case the spatial (wavenumber) Eqs. (13) and (16) should be replaced by temporal (frequency) equation
 $$
 E(f)   \propto \int_0^{\infty} P(f_c) \exp -(f/f_c)~ df_c  \propto \exp-(f/f_{\beta})^{\beta},   \eqno{(23)}
 $$

    Instead of the  Eqs. (18) and (19) one obtains from the dimensional considerations 
$$
 u_c \propto |I_n|^{1/(4n-3)}~ f_c^{\alpha_n}    \eqno{(24)}
 $$    
where 
$$
\alpha_n = \frac{2n-3}{4n-3}  \eqno{(25)}
$$  
and instead of the Eq. (21) 
$$
 \beta_n = \frac{2(2n-3)}{(8n-9)}   \eqno{(26)}  
 $$

\begin{figure} \vspace{-1.57cm}\centering
\epsfig{width=.45\textwidth,file=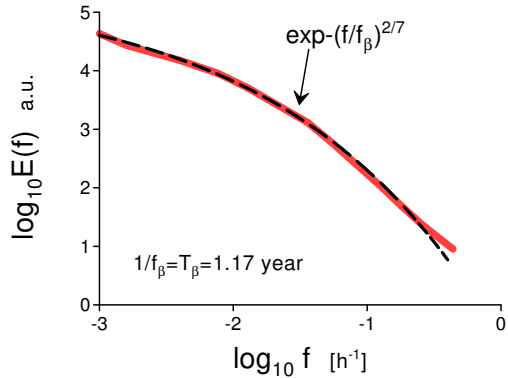} \vspace{-3.7cm}
\caption{Power spectrum of the atmospheric $CO_2$ temporal fluctuations - low latitude (northern hemisphere). } 
\end{figure}
\begin{figure} \vspace{-0.5cm}\centering
\epsfig{width=.45\textwidth,file=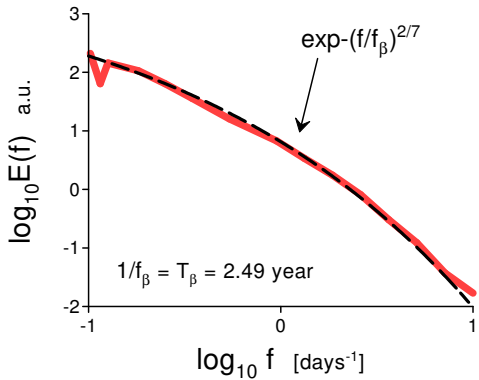} \vspace{-3.5cm}
\caption{Power spectrum of the atmospheric radon ( $^{222}Rn$) temporal fluctuations - middle latitude (southern hemisphere). } 
\end{figure}

  Then for the strongly chaotic helical distributed chaos ($n=2$) one obtains $\beta = 2/7$, i.e.
  
$$
E(f) \propto \exp-(f/f_{\beta})^{2/7}  \eqno{(27)}
$$ 
for the proper temporal (frequency) spectrum. \\

  For $n=3$ one obtains $\beta = 2/5$, i.e.
  
$$
E(f) \propto \exp-(f/f_{\beta})^{2/5}  \eqno{(28)}
$$ 
for the proper temporal (frequency) spectrum. \\

  Figure 13 shows power spectrum of the atmospheric nitrogen dioxide $NO_2$ {\it temporal} fluctuations measured in the center of the Moscow megacity (atmospheric surface layer). The pronounced peaks at 24, 12 and 8 hours correspond to the diurnal periodicity and its harmonics. The spectral data were taken from the Fig. 6 of the Ref. \cite{gor}. The dashed curve in the Fig. 13 indicates correspondence to the spectrum Eq. (28).

  Figures 14-16 show power spectra of the atmospheric $CO_2$ temporal fluctuations measured in the atmospheric surface layer at the meteorological stations Barrow, USA (high latitude), Schauinsland, Germany (middle latitude), and Hateruma, Japan (low lattidude), respectively (all three stations are located in the northern hemisphere). The spectral data were taken from the Fig. 2b of the Ref. \cite{patra}. The peaks in the Figs. 14 and 15 correspond to the diurnal (24 hours) periodic variability. The dashed curves in the Figs. 14-16 indicate correspondence to the spectrum Eq. (27).\\
    
   Figure 17 shows power spectrum of the atmospheric radon ( $^{222}Rn$) temporal fluctuations measured in the atmospheric surface layer at the Cape Point station (South Africa). This is a middle latitude coastal station in the Southern Hemisphere. The spectral data were taken from the Fig. 4 of the Ref. \cite{both}. The dashed curve in the Fig. 17 indicates correspondence to the spectrum Eq. (27). It should be noted that $T_{\beta}$ are about the same ($\simeq 2.5$ year) for the {\it middle} latitude northern and southern hemispheres' observations (cf. Figs. 15 and 17, see also Ref. \cite{gb} and references therein).\\

 \section{Solar radiation transmittance}

   Properties of the atmospheric solar radiation transmittance are closely related to the properties of the atmospheric tracers, especially in presence of the clouds. The optical properties of the clouds are determined by their composition and its chaotic/turbulent dynamics. Generally, the droplets, aerosols etc. cannot be considered as dynamically passive scalars in this case. However, taking into account the above suggested separation of the scales for the helicity effects one can expect that the above consideration can be (with certain restrictions) applied to this case as well: i.e the effects related to the active role of the droplets, aerosols etc. are non-relevant to the inertial range of scales. \\
   
\begin{figure} \vspace{-1.1cm}\centering
\epsfig{width=.45\textwidth,file=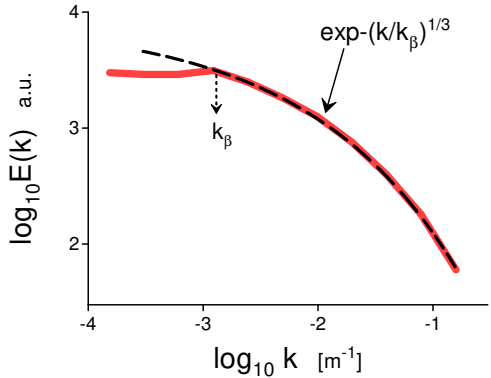} \vspace{-3.85cm}
\caption{Power spectrum of the global radiation transmittance (vs. the horizontal wavenumbers $k$) for the broken-cloud sky conditions.} 
\end{figure}
\begin{figure} \vspace{-0.3cm}\centering
\epsfig{width=.45\textwidth,file=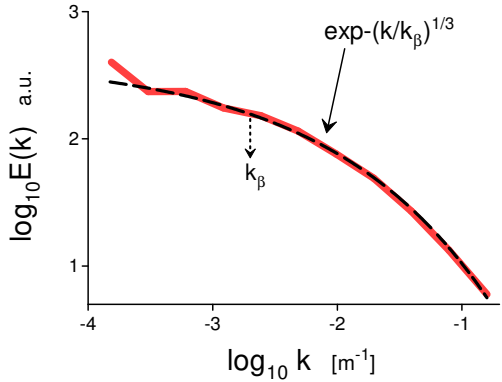} \vspace{-4.2cm}
\caption{As in the Fig. 18 but for the cirrus sky conditions. } 
\end{figure}
\begin{figure} \vspace{-1.1cm}\centering
\epsfig{width=.45\textwidth,file=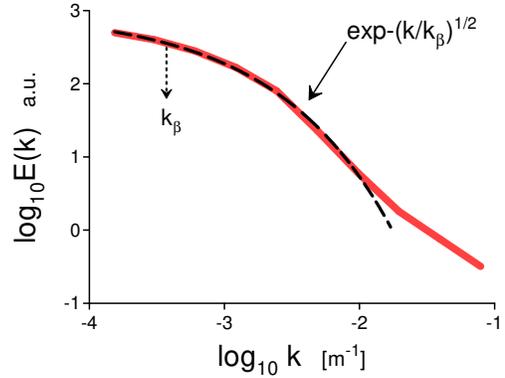} \vspace{-3.85cm}
\caption{As in the Fig. 18 but for the overcast cloudy sky conditions.}
\end{figure}
     
   In the recent paper Ref. \cite{madh} results of measurements of the global radiation, performed using a ground-based (surface) dense network of 99 pyranometers for different sky conditions, were reported. The global transmittance was computed by normalizing the global radiation by the extraterrestrial radiation at the top of atmosphere and by accounting for the Sun–Earth distance and for the cosine of solar zenith angle. The experimental site ($10 \times 12$ km) was located near Julich (Germany) and the measurements were produced 
 from April 2 to July 24, 2013 during the daylight periods. Transformation from frequency to wavenumber space was made using the Taylor ``frozen'' turbulence hypothesis: i.e. assuming that  the variability in the measured surface radiation was dominated by the advection of the spatial structures of the cloud fields across the local point of the observation rather than their temporal local  variability (see above).\\

    Figures 18-20 show power spectra of the global transmittance (vs. the horizontal wavenumbers $k$) for the broken-cloud, cirrus and overcast sky conditions respectively. The spectral data were taken from the Fig. 5 of the Ref. \cite{madh}. The dashed curves in the Figs. 18, 19 indicate correspondence to the spectrum Eq. (22) whereas the dashed curve in the Fig. 20 (overcast cloudy sky) indicates correspondence to the spectrum Eq. (15).\\
    
    One can see that the power spectra of the solar radiation transmittance computed for the data obtained in the cloudy atmosphere correspond to those predicted (and observed) for the passive scalar tracers.

\end{document}